\documentclass[preprintnumbers,aps, pre,twocolumn, floatfix, showpacs, showkeys]{revtex4-1}

\pdfoutput=1
\usepackage{graphicx}
\usepackage{dcolumn}
\usepackage{tabularx}
\usepackage{bm}
\usepackage{color}

\begin{document}

\bibliographystyle{apsrev}

\title{Flow Motifs Reveal Limitations of the Static Framework \\to Represent Human interactions}

\author{Luis E C Rocha}\email{Luis.Rocha@uclouvain.be}
\author{Vincent D Blondel}

\affiliation{Department of Mathematical Engineering \\
Catholic University of Louvain \\
Louvain-la-Neuve, Belgium}

\date{\today }

\begin{abstract}

Networks are commonly used to define underlying interaction structures where infections, information, or other quantities may spread. Although the standard approach has been to aggregate all links into a static structure, some studies suggest that the time order in which the links are established may alter the dynamics of spreading. In this paper, we study the impact of the time ordering in the limits of flow on various empirical temporal networks. By using a random walk dynamics, we estimate the flow on links and convert the original undirected network (temporal and static) into a directed flow network. We then introduce the concept of flow motifs and quantify the divergence in the representativity of motifs when using the temporal and static frameworks. We find that the regularity of contacts and persistence of vertices (common in email communication and face-to-face interactions) result on little differences in the limits of flow for both frameworks. On the other hand, in the case of communication within a dating site (and of a sexual network), the flow between vertices changes significantly in the temporal framework such that the static approximation poorly represents the structure of contacts. We have also observed that cliques with 3 and 4 vertices containing only low-flow links are more represented than the same cliques with all high-flow links. The representativity of these low-flow cliques is higher in the temporal framework. Our results suggest that the flow between vertices connected in cliques depend on the topological context in which they are placed and in the time sequence in which the links are established. The structure of the clique alone does not completely characterize the potential of flow between the vertices.

\end{abstract}

\pacs{89.75.Hc, 89.75.Fb, 02.50.Ey}
\keywords{temporal networks, flow graphs, motifs}
\maketitle

\section{Introduction}

Networks have been used to describe and model natural and artificial systems from different scientific domains. The common practice has been to construct a network by aggregating all available information into a single graph structure, eventually containing weights or directions on the links to reflect attributes of the system~\cite{Newman10, Costa11}. The main underlying assumption is that links remain fixed during the sampling interval such that all information of the system is  summarized into the network topology. Although static structures provide several insights about the patterns of interaction, neglecting the timing of events results in an incomplete picture of the interaction dynamics. The time order of events may reveal cyclic behavior~\cite{Holme03, Isella11}, bursts of activity~\cite{Holme03, Barabasi05, Rocha10}, causality~\cite{Kovanen11, Lentz12}, and long-term correlations~\cite{Rybski09, Rocha10}. One typical illustration of the ordering effect is the absence of paths between vertices A and C if B and C are linked before A and B. In the static version however this path would exist either way. The temporal constrains significantly reduce the number of paths and increase the path-length between two vertices~\cite{Tang10, Pan11}, directly affecting the centrality of vertices~\cite{Grindrod11, Takaguchi12, Konschake13}.

\begin{figure}[htb]
\centering
\includegraphics[scale=0.35]{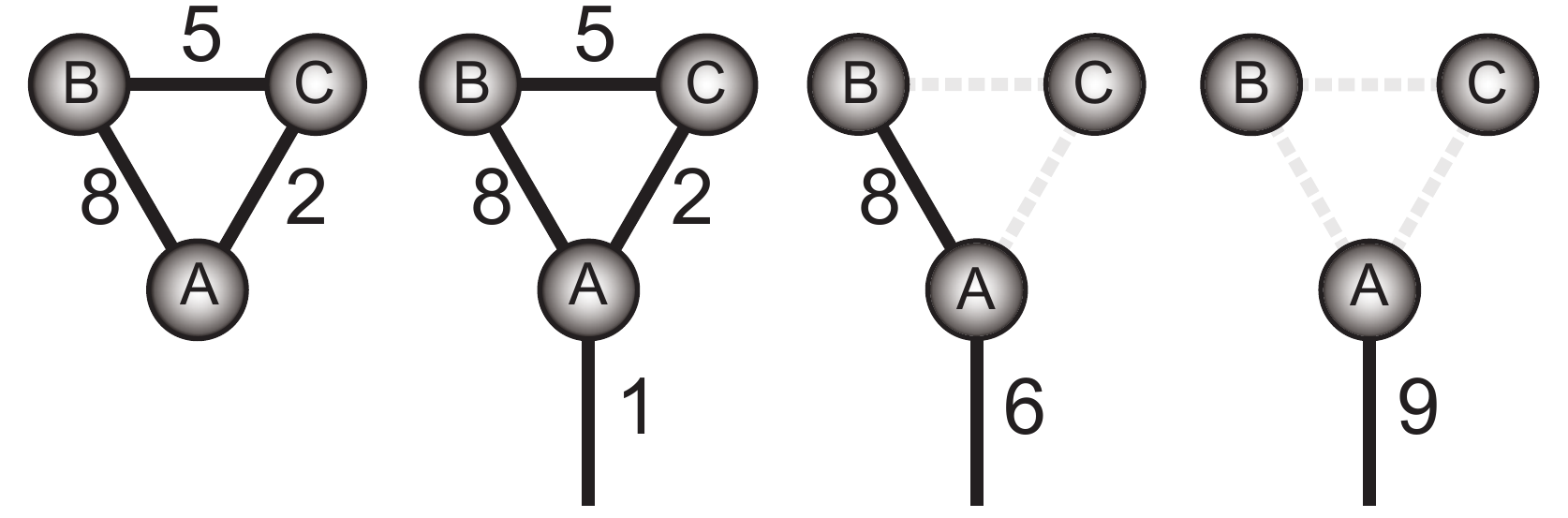}
\caption{Sample of a temporal network. The paths between pairs of vertices change according to the time stamps in the links.}
\label{panel_1}       
\end{figure}

The network structure defines potential pathways for the transmission of infection, information, or other quantities. In the context of dynamical processes, several studies suggest that temporal patterns of link activity reflect into the spread by speeding up or slowing down epidemics~\cite{Rocha13, Takaguchi12b, Rocha11, Stehle11}, information~\cite{Miritello11, Karsai11, Lionel12}, random walks~\cite{Starnini12, Haerter12, Perra12}, and consensus dynamics~\cite{Takaguchi11}. Most of the studies about dynamics on temporal networks focus on global statistics and on detecting the timing for emergent phenomena. There is little understanding, however, on local effects of the time ordering. Consider for example the triangles on Figure~\ref{panel_1}. If vertices A and C connect (at $t=2$) before B and C ($t=5$), B may infect A directly (at $t=8$), but not through C. Paths get more complicated when another link is added to A; depending on the time-stamp of the new link, different routes are available for an infection to reach B and C (if $t=1$), or to not reach them at all (if $t=9$). In temporally large networks (i.e.\ several days, or weeks), a particular sequence may not affect much the dynamics because links are likely to repeat, e.g.\ the time order of events in Fig.~\ref{panel_1} may repeat every day and an infection would pass from B to A in one day and then to C at the day after. Although intuitively one may expect that the infection or information flowing between adjacent vertices always change if time constrains are introduced, these changes in fact depend on the particularities of the system.

In this paper, we estimate the limits of flow between vertices in empirical temporal networks and compare with the respective static version. To do that, we let a random walker explore the network (both the temporal and the static versions) and measure the frequency of hopes between two vertices to estimate the local flow. Since the walker moves according to the availability of links in the temporal network, the flow captures the interplay between temporal and topological constrains. In the static network however links are available at any moment and thus only the topology affects the dynamics. This procedure converts the original simple network into a flow network containing information about the dynamics on the links. The concept of flow networks has been exploited before, in the context of static networks, to characterize the community structure~\cite{Rosvall08, Kim10} and the spreading dynamics~\cite{Lambiotte11}. Since this measure of flow is globally optimized, in order to quantify relative changes between the temporal and the static frameworks, we split the links into two classes of high- and low-flow. To quantify the impact of temporal constrains in the flow, we measure the representativity of flow sub-graphs (or flow motifs) in temporal networks and compare with the static version. In this context, motifs are convenient because they summarize the information about the system-wide flow into small subunits that can be studied individually but collectively provide a fingerprint of the network~\cite{Milo02, Milo04, Alon06, Knabe08, Costa08, Sinatra10, Kovanen11}.

\section{Materials and Methods}

In this section, we describe the empirical networks used in this paper, the procedure to estimate the flow using the random walker dynamics on temporal and static networks, and the methodology to obtain flow sub-graphs and flow motifs.

\subsection{Temporal networks}

A temporal network of size $N$ is defined as an ordered sequence of triplets $(i,j,t)$. Each triplet corresponds to a single interaction event at time $t$ ($t=0,1,2,...$) between a pair $(i,j)$ of vertices $i$ and $j$ ($i,j=1,2,...,N$). The total number of triplets (or the number of contacts or single event interactions) is given by $C$ and the number of unique pairs $(i,j)$ (i.e.\ links in the static network) is given by $E$. An empirical temporal network is a set of interaction events recorded every $\delta t$ (the sampling resolution) during an interval $T_{\text{i}} \le t \le T_{\text{f}}$. The sample interval $\Delta T$ is thus $T_{\text{f}} - T_{\text{i}}$. The corresponding static network is obtained by aggregating all interactions events within $\Delta T$ into a single graph structure and using the number of contacts per link as weights.

We use a selection of empirical temporal networks to represent different classes of human communication and proximity contacts. These networks represent diverse underlying structures relevant to the spread of information and of infections. The first dataset corresponds to a network of online communication between members of a dating site (POK)~\cite{Holme04}. The second corresponds to email communication within a major university (EMA)~\cite{Eckmann04}. The third corresponds to sexual contacts between sex-buyers and -sellers (SEX) and is fully bi-partite by definition ~\cite{Rocha10}. The other datasets correspond to face-to-face interactions between delegates in a scientific conference (SPC) and to visitors of a museum exposition (SPM)~\cite{Isella11}. All networks are assumed undirected. Table~\ref{tab_1} summarizes the fundamental characteristics of the network samples used hereafter.

\begin{table}[htb]
\centering
\begin{tabular}{cccccc}
\hline
 & $N$ & $C$ & $E$ & $\delta t$ & $\Delta T$ (day) \\
\hline
POK & $13,225$   & $167,909$     & $40,821$   & $1$ min    & $100$    \\
EMA & $3,180$    & $301,815$     & $31,664$   & $1$ min    & $80$     \\
SEX & $12,157$   & $34,060$      & $26,383$   & $1$ day    & $1,000$  \\
SPC & $113$      & $20,818$      & $2,196$    & $20$ sec   & $2.5$    \\
SPM & $72$       & $6,980$       &  $691$     & $20$ sec   & $1$      \\
\hline
\end{tabular}
\caption{Characteristics of the samples of the empirical networks: number of vertices ($N$), number of single event interactions ($C$), number of unique pairs of connected vertices ($E$), the network resolution ($\delta t$), and the sample interval $\Delta T$.}
\label{tab_1}
\end{table}

\subsection{Random walker dynamics}

The random walk is the simplest dynamic process one may study in a network and conveniently gives a measure of flow within the network~\cite{, Rosvall08, Kim10, Lambiotte11, Lambiotte12}. In a time ordered network, the concept of flow given by the random walk dynamics remains intuitive but the implementation of the process is not as simple as in the static network because time introduces a few new constrains. In particular, the dataset is limited within a time interval and thus border effects appear in the beginning and ending of the interval, the interactions are sparse and thus the walker may get trapped in a vertex for a while, and at last, there is a time scale.

In our model, the walker starts its network tour in a uniformly chosen vertex and only moves forward (jump or hop) whenever there is a contact active, choosing uniformly if more than one contact is available (e.g.\ a vertex contacts two other vertices at the same time). The walker performs a large number of hops ($H \ge 10^7 \gg C$) in order to explore as much as possible the network and reach the stationary state. We count the number of hops $h_{\text{ij}}$ from vertex $i$ to $j$ and normalize by the total number of hops $H$ to obtain an estimative of the flow from $i$ to $j$, i.e.\ $f_{\text{ij}} = h_{\text{ij}}/H$. Note that for the static network, the stationary solution of the flow in the links can be calculated directly from a modified version of the google matrix (see ref.~\cite{Kim10}).

For simplicity, the walker step corresponds to one time unit (which is set as the temporal resolution). After each hop, time is incremented. In static networks, there is always the possibility to move except in case of isolated vertices. On the other hand, in temporal networks the walker may have to wait in a vertex until a contact becomes available. Since this waiting time is not accounted, the walker dynamics is not affected in case of absence of contacts during some periods of time (e.g.\ over night) because the walker remains still, in the vertex, during the period of inactivity. Nevertheless, if the vertex has only one contact, the walker may get trapped in that vertex because after the arrival it cannot move forward. To avoid such situation, at each time step, the walker at vertex $i$ teleports to a uniformly chosen vertex $j$ ($i\neq j$) at a uniformly chosen time $t$ ($T_{\text{i}} \le t \le T_{\text{f}}$) with probability $\tau$ or hops to an adjacent vertex with probability $1-\tau$. If the walker reaches the last time step, it teleports with $\tau=1$. The teleportation makes the walk ergodic and is not counted in the flow to improve its estimative~\cite{Lambiotte12}.

\begin{figure}[htb]
\centering
\includegraphics[scale=1]{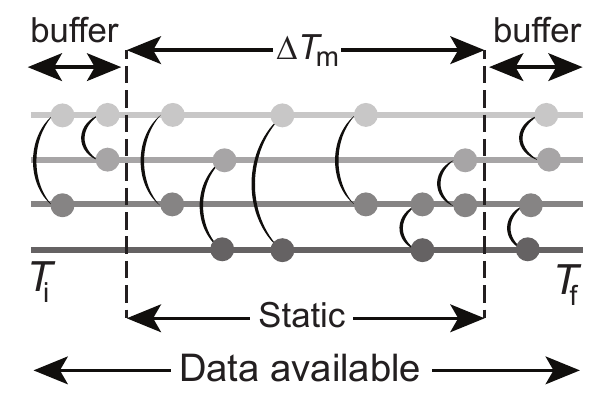}
\caption{The sample of the empirical network is available between times $T_{\text{i}}$ and $T_{\text{f}}$. The random walker moves in, out, and within the buffer regions but only hops within $\Delta T_{\text{m}}$ are counted in the flow. The respective static network contains only the links made within $\Delta T_{\text{m}}$.}
\label{panel_2}
\end{figure}

To remove the finite size effects due to the limited sampling window, we adopt open boundary conditions. This is done by defining two buffer intervals given respectively by $[T_{\text{i}} , T_{\text{i}} + T_{\text{buffer}}]$ and $[T_{\text{f}} - T_{\text{buffer}}, T_{\text{f}}]$, i.e.\ one in the beginning and one in the end of the network (Fig.~\ref{panel_2}). The walker may move in, out, and within these regions but only hops between $(T_{\text{i}} + T_{\text{buffer}}) < t < (T_{\text{f}} - T_{\text{buffer}})$ are used to estimate the flow. The length of the buffer is $T_{\text{buffer}}=1.5/\tau$ time steps (with contacts). We use $\tau=0.1$, which implies on $15$ time steps lost on each side. This choice guarantees that a walker reaching the end of the interval $\Delta T_{\text{m}}$ continues moving (within the buffer and thus without the interference of the border) until it teleports due to $\tau$ (one expects on average $1/\tau$ hops before teleportation, thus $1.5/\tau$ is chosen to account for the fluctuations). The same reasoning applies on the left border of the network. Such open boundaries imply that the walker does not see the borders of the network. One can intuitively understand the meaning of open boundaries by imagining an infinite system where the walker can hop everywhere but only hops within the interval $\Delta T_{\text{m}}$ are counted in the statistics (Fig.~\ref{panel_2}).

By using periodic boundary conditions, we would not compromise any links (in the buffer) but we introduce undesired biases due to the artificial periodicity. For example, imagine that vertex B is connected to A and C; if vertex A and B interact before B and C, the walker can move from A to B, and from B to C in two hops, but if the links are artificially periodic, the walker may move back from C to B and then, from B to A, in two hops as well. We expect that such biases are more evident in temporally sparse networks. A fine analysis of the impact of boundary conditions is out of the scope of our study and we therefore simply adopt the open boundaries for the reasons stated above. Finally, the static version of the empirical network is thus obtained by collecting only the links made within the interval $\Delta T_{\text{m}}$ and is thus weighted.

\subsection{Flow motifs}
\label{sec_flow}   

By coupling the estimated flow to the links, we convert the original network structure into a flow network that simultaneously contain information about the topology and about the diffusion dynamics. In order to compare the limits of flow due to the time ordering with the absence of time, we take a coarse grained approach and split the links into two categories of high- and low-flow. The splitting point is given by the median of the estimated flow such that an equal number of links are labeled red (for high-flow) and blue (for low-flow). This procedure defines high- and low-flow relatively to all links in the same network and independently of the framework adopted (temporal or static). Since the flow is not necessarily symmetric between the same pair of vertices, the flow network has directed links.

\begin{figure}[tb]
\centering
\includegraphics[scale=0.5]{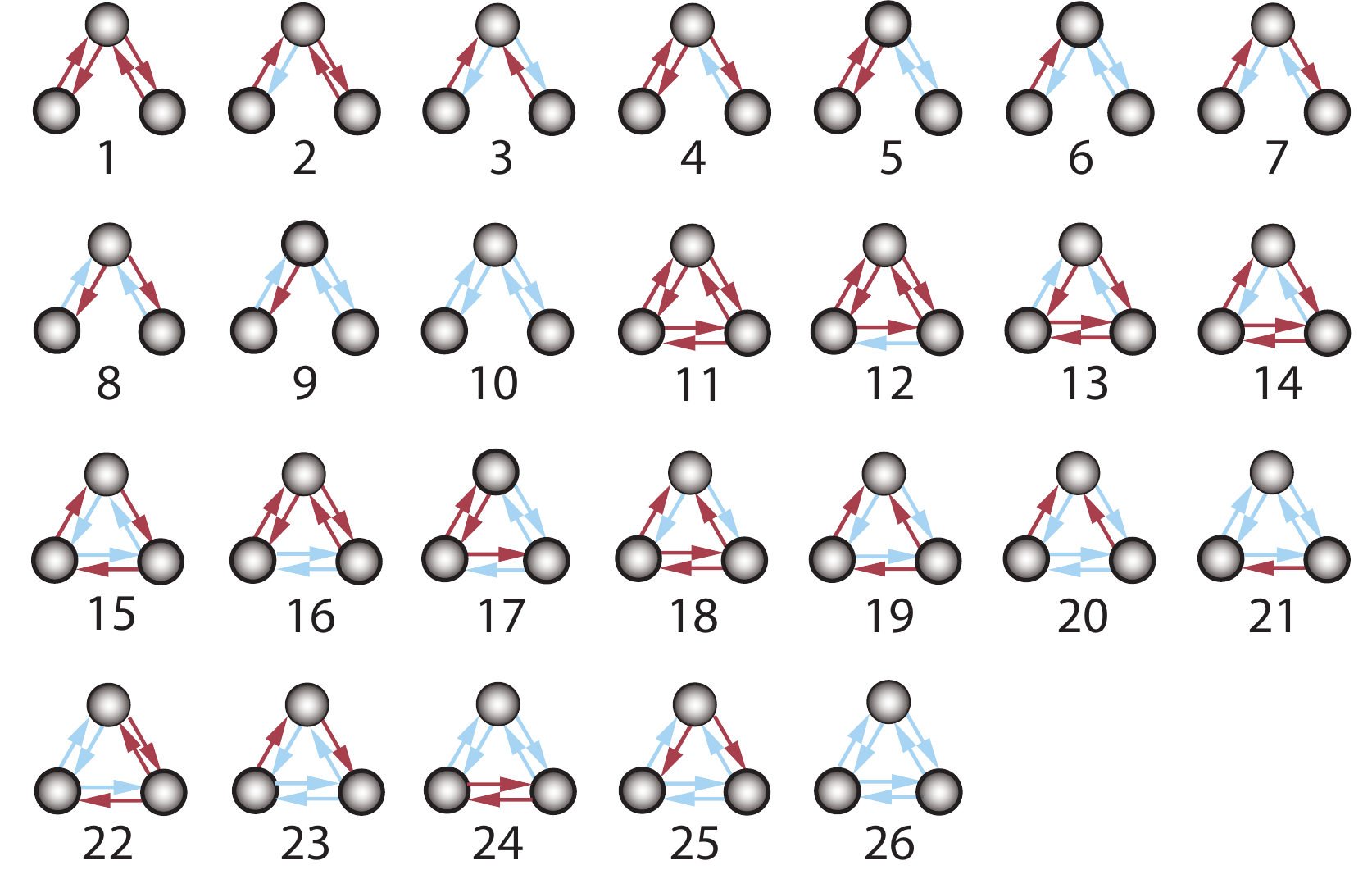}
\caption{All $26$ types of connected triads, i.e.\ sub-graphs with $3$ vertices. There are $10$ triplets (sub-graphs $1$ to $10$) and $16$ triangles (sub-graphs $11$ to $26$). Red links correspond to high-flow and blue links to low-flow. Bipartite networks only have triplets.}
\label{panel_3}       
\end{figure}

To study the flow network, we analyze its various motifs. Motifs (anti-motifs) are sub-graphs containing a number of vertices and combinations of connected pairs of vertices, observed more (less) often than one expects by chance~\cite{Alon06}. In some contexts, motifs are believed to correspond to single regulatory units designed to perform specific functions and tasks~\cite{Milo02, Milo04, Alon06, Knabe08}, while in other contexts, they are used to identify representative sequences of events~\cite{Sinatra10, Kovanen11}. All together, motifs create a fingerprint of the network~\cite{Milo04}. We define a flow motif by measuring the statistical representativity of the sub-graphs. This is done by comparing the frequency of the sub-graph in the flow network ($\mu_{\text{ori}}$) and in randomized versions of the same network. The randomized networks are generated by uniformly selecting two links and switching one of the contacts of each link. The in- and out-degree, the bi-partivity (if applicable), and the labels of the vertex are all conserved~\cite{Wernicke06}. The representativity is thus given by the Z-score, i.e.\ $Z=(\mu_{\text{ori}} - \langle \mu_{\text{rand}} \rangle)/\sigma$, where $\langle \mu_{\text{rand}} \rangle$ is the average frequency of the sub-graph in the random ensembles and $\sigma$ is the respective standard deviation~\cite{Milo02, Alon06}. To illustrate the concept of flow networks, Figure~\ref{panel_3} shows all $26$ types of $3$-vertex flow sub-graphs.

\section{Results}

In this section, we compare the flow properties of temporal networks in respect to the static network, and discuss the characteristics and importance of the most abundant flow motifs.

\subsection{Deviation from the static network}

We start our analysis by counting the number of links with the same label (high- and low-flow, or red and blue) in both temporal and static networks. We identify that the proportion of links similar in both frameworks varies according to the category of the network, the lowest being the SEX network with $63.1\%$ and the highest being the SPM network with $86.6\%$. This means that about $37\%$ and $13\%$ of the links change the category of flow intensity (high to low, or low to high) if the time information is included. This discrepancy is understandable since the SEX network is temporally sparse and vertices are mostly active during relatively short intervals~\cite{Rocha10}, in such scenario, the time order plays an important role because cycles or repetitions of contacts are unlikely. In the SPM network, vertices correspond to individuals visiting a museum in well defined time-slots~\cite{Isella11}, which means that a group of vertices remains well connected during the visit and loosely connected with the group of the next (or previous) time-slot. In such case, one expects that the static network closely represents the real interaction dynamics. The other cases contain intermediate results consistent with the category of the networks, i.e.\ $72.2\%$ (POK), $74.4\%$ (SPC), and $79.13\%$ (EMA).

The proportion of links with same label in both frameworks indicates that, to different extends, time order alters the flow on the links on each category of network. One may expect that the matching of colors on both frameworks decreases if more than one link is compared simultaneously, e.g.\ the label of all links connected to a vertex. Here, however, we want to study how the system-wide characteristics of the flow change from one framework to another. Therefore, we look into the statistics of the sub-graphs. To simplify our analysis, we focus on two categories of networks, POK and EMA, corresponding to the largest datasets available (Table~\ref{tab_1}) and provide summary statistics for the other networks. The networks POK and EMA have the same resolution $\delta t$ and approximately the same $\Delta T$ (Table~\ref{tab_1}).

\begin{figure}[htb]
\centering
\includegraphics[scale=1.5]{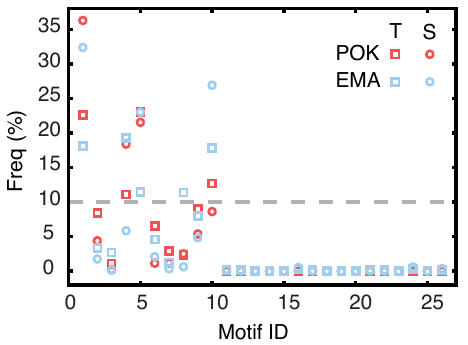}
\caption{Relative frequency of the sub-graphs in the temporal (T) and static (S) networks for the POK and EMA networks.}
\label{panel_4}       
\end{figure}

If we count the frequency of the sub-graphs in the networks POK and EMA, we identify that triplets (sub-graphs $1$ to $10$, see Fig.~\ref{panel_3}) are in general significantly more common than triangles irrespective of the framework adopted (Fig.~\ref{panel_4}). This simply reflects the sparsity of the networks. The four sub-graphs that together account for more than $60\%$ of the occurrences are, in order of decreasing frequency, sub-graphs $1$, $10$, $5$ and $4$. With the exception of sub-graph $4$, the other sub-graphs reflect symmetric relations between the central vertex and the two contacts. Sub-graph $4$ typically represents the periphery of the network where one of the leaf vertices receives high-flow but transmits low-flow. In both networks, the frequency of sub-graph $1$ decreases by about $40\%$ in the temporal version in comparison to the static case. Sub-graph $10$ also decreases by the same amount in the EMA network but increases its frequency in the case of POK.

\begin{figure}[htb]
\centering
\includegraphics[scale=1]{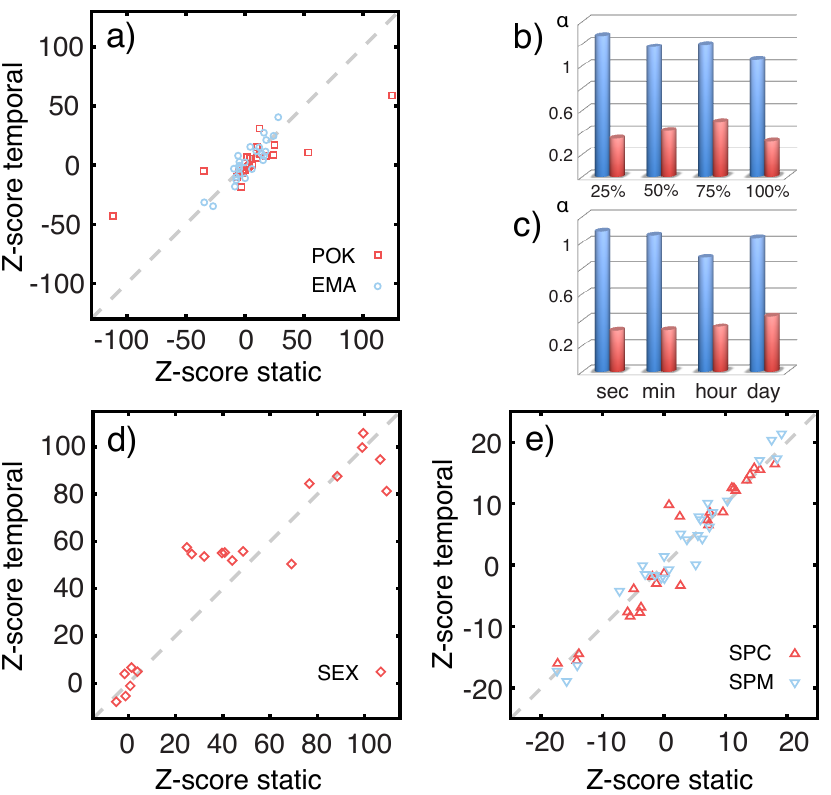}
\caption{a,d,e) Relation between the Z-scores of the $3$-vertex sub-graphs for the static and for the temporal version of the same network. Dashed line corresponds to $\alpha=1$ in $f(x)=\alpha x + \beta$; b) The slope $\alpha$ of the linear regression between the Z-scores in the static and in the temporal frameworks for POK (red) and EMA (blue). Each group corresponds to different fractions of the original network (given in percent); c) The slope $\alpha$ of the linear regression for various temporal resolutions.}
\label{panel_5}
\end{figure}

Although the frequency of sub-graphs illustrates some general properties, the Z-score provides a robust measure of significance, since in this case, the frequency of the sub-graph is compared to what one expects by chance. If we compare the Z-score of the $3$-vertex sub-graphs detected in the temporal network and in the respective static version, we observe a positive correlation between the two cases (Fig.~\ref{panel_5} and Table~\ref{tab_2}).  By fitting a straight line ($f(x)= \alpha x+\beta$, where $x$ is the Z-score static and $f(x)$ is the Z-score temporal), we see that the results for POK network deviate significantly from $\alpha=1$ (dashed line) while in the EMA network, $\alpha = 1.06$ (Fig.~\ref{panel_5}a). In the case of POK, highly represented motifs (and anti-motifs) in the static network are significantly less represented in the temporal version ($\alpha = 0.32$). On the other hand, for EMA, motifs (and anti-motifs) with large absolute Z-score in the static network, which is the reference network, are only slightly more represented in the temporal network. These linear relation characterized by the slope $\alpha$ is robust to variations in the sample size of the networks (Fig.~\ref{panel_5}b) as well as to different temporal resolutions (Fig.~\ref{panel_5}c). For decreasing resolution, the temporal network should converge to the static case. We see that tuning the resolution from seconds to day, $\alpha$ slowly increases for the POK network. The results suggest that the discrepancy between the temporal and static frameworks is indeed characteristic of these systems. The SPC and SPM networks have similar characteristics to EMA, but SEX has the same tendency as the POK network with a relatively higher $\alpha$ (Fig.~\ref{panel_5}d,e and Table~\ref{tab_2}). In the SEX case, however, a number of sub-graphs (with Z-scores static between $20$ and $70$), corresponding to highly asymmetric combinations (e.g.\ sub-graphs $2, 4, 6$, and $9$ in Fig.~\ref{panel_3}), is unrelated in the static and in the temporal frameworks.

\begin{table}[htb]
\centering
\begin{tabular}{cccc}
\hline
         & $\alpha$             & $R^2$     & p-value \\
\hline
POK  & $0.32 \pm 0.05$ & $0.640$  & $<0.01$ \\
SEX   & $0.86 \pm 0.08$ & $0.867$  & $<0.01$ \\
SPC & $1.04 \pm 0.06$ & $0.930$  & $<0.01$ \\
SPM  & $1.05 \pm 0.04$ & $0.958$  & $<0.01$ \\
EMA  & $1.06 \pm 0.07$ & $0.897$  & $<0.01$ \\
\hline
\end{tabular}
\caption{Statistics of the relation between the temporal and static frameworks. The slope $\alpha$ of the linear regression between the Z-scores in the static and in the temporal frameworks. The square of the sample correlation $R^2$ and the respective p-value.}
\label{tab_2}
\end{table}

The POK and EMA networks represent two distinct classes of temporal networks; in the POK network, the vertex turnover is much higher than in the EMA network. If we measure the fraction of vertices both in the initial and in the final $5\%$ of active links, we identify that only $1.5\%$ of the vertices belong to this group in the POK network (and $1.7\%$ in the SEX) while this number is $47.9\%$ for the EMA network (and $38.9\%$ in the SPC). A high vertex turnover rate means that vertices enter and leave the system frequently such that contacts (or partnerships) vary in time.  This is not surprising since in the dating site (POK), members are active for some period and then leave the community either because they found a date or because they are consistently unsuccessful (similar dynamics happen in the network of sex-buyers and -sellers, where there is a strong tendency to not repeat partners and to not stay long in the system~\cite{Rocha10}). As a consequence, the collapse of the temporal network into a static structure mixes links between vertices made at different moments. In the context of spreading processes, this mixing homogenizes the flow between close vertices because it removes the time constrain. On the other hand, low turnover is associated to increased network stability in the sense that individuals tend to interact with roughly the same partners in time. This is expected in the email exchange context (EMA) where individuals tend to interact with about the same people within a period of few months (or during a conference~\cite{Isella11}). Due to the regularity of the contacts, the transmission potential of a particular link is not significantly affected. The SPM consists of a particular case because, within the same time slot, individuals interact similarly to EMA or SPC cases (i.e.\ adjacent vertices remain the same and interactions are persistent) and only few links connect visitors of different time slots (in the static framework, this effect is reflected into a strong community structure~\cite{Isella11}). In other words, the static framework is adequate to estimate the limits of flow in the case of regularity and stability of contacts, typically with low vertex turnover; on the other hand, the static framework poorly approximates the contact patterns in case of high vertex turnover and temporal variation of partnerships.

\subsection{Individual flow motifs}

In this part, we compare the representativity of the flow sub-graphs for two classes of networks represented by POK and EMA networks. In the context of static directed networks, it is known that the representativity of each sub-graph varies for different categories of networks. For example, the feed-forward motif is commonly observed in transcription networks but poorly represented in the networks of adjacent words in written texts~\cite{Milo04}. In order to compare networks of different sizes, it is necessary to normalize the vector of Z-scores to length $1$ by using $NZ_i=Z_i/(\sum_j Z_j^2 )^{0.5}$. Motifs in large networks display higher Z-scores than in small networks~\cite{Milo02}; the normalization corrects that by measuring the relative significance of sub-graphs.

\begin{figure}[tbh]
\centering
\includegraphics[scale=0.5]{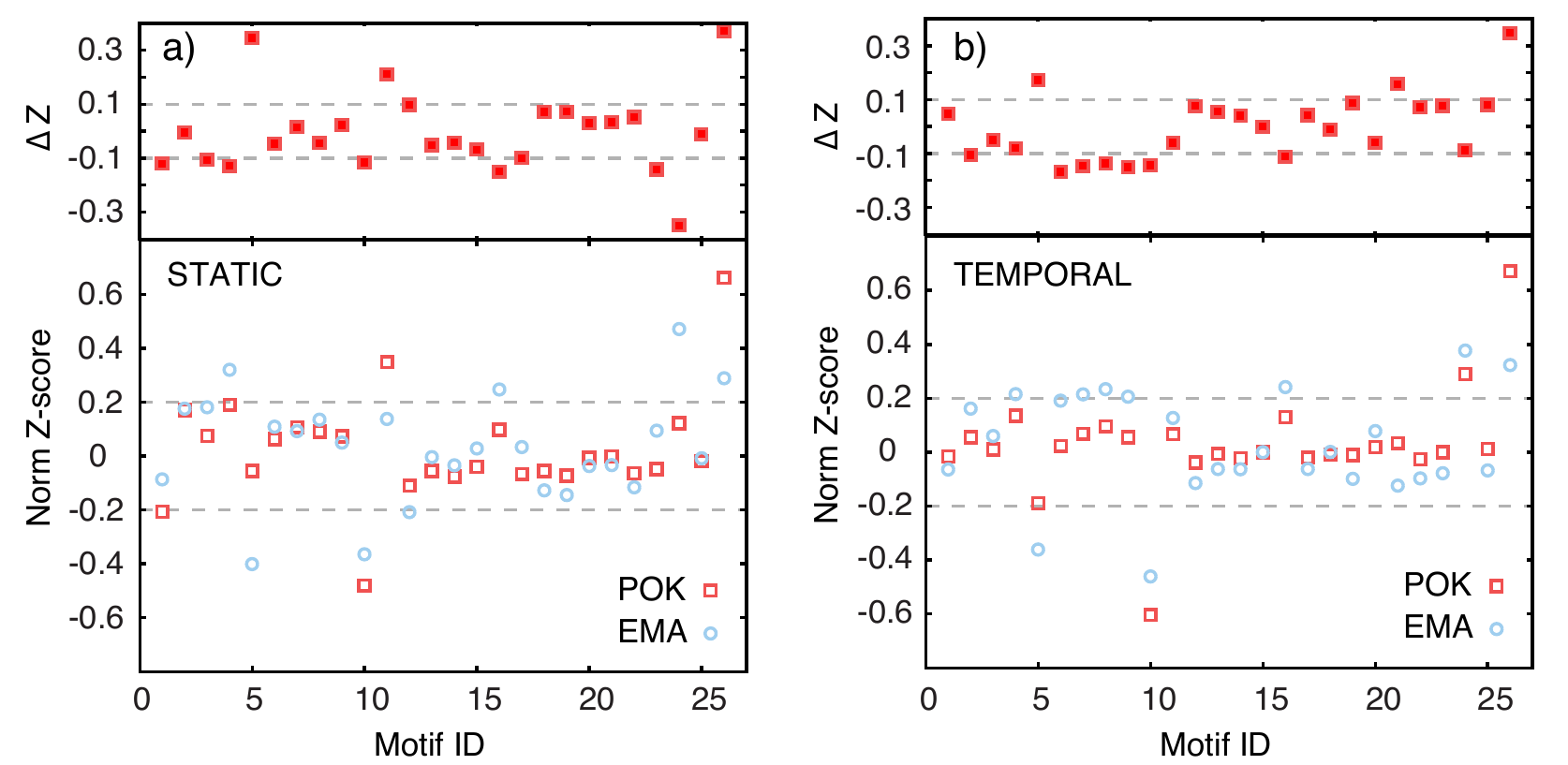}
\caption{a) Normalized Z-scores of the sub-graphs for the static networks; b) Normalized Z-scores of the sub-graphs for the temporal networks. On top of a) and b), the difference $\Delta Z = NZ_{\text{POK}}-NZ_{\text{EMA}}$ is presented, therefore, $\Delta Z>0$ means that a particular sub-graph is more abundant in the POK than in the EMA network.}
\label{panel_6}
\end{figure}

Figure~\ref{panel_6}a (bottom) shows the normalized Z-scores for the static POK and EMA networks. Since there is no threshold to define a motif or anti-motif, the dashed lines at $\pm 0.2$ are simply reference lines. In the POK network, sub-graphs $26$ and $11$ are respectively the most represented motifs, and sub-graph $10$ is the most anti-motif. On the other hand, in the EMA network, sub-graphs $24, 4$, and $26$ are the most represented motifs, and sub-graphs $5$ and $10$ are anti-motifs. This reflects both the clustering structure (since motifs $11, 24$ and $26$ are triangles) and the difference in the contact patterns in both networks. In fact, taking the difference of the normalized Z-scores between the POK and EMA networks (Fig.~\ref{panel_6}a (top)), we identify that motifs $24$ and $26$, and anti-motif $5$ are significantly different in both networks.

In the case of temporal networks, the representativity of sub-graphs is more distinguishable between both networks (Fig.~\ref{panel_6}b). Most sub-graphs are poorly represented in the POK network (NZ-score close to zero), but in the case of EMA, the triplets are typically well represented (NZ-score close to $0.2$) while triangles are under represented (NZ-score close to $-0.1$). In the temporal framework, for both networks, sub-graphs $24$ and $26$ are motifs and sub-graphs $5$ and $10$ are anti-motifs. The largest differences are observed respectively for sub-graphs $26$ and $5$.

These observations indicate that triangle structures (i.e.\ the fact that two friends are also friends themselves) do not necessarily indicate high flow between the vertices in the context of temporal networks. In fact, in a triangle, it is more likely to find low flow between all three pairs of vertices (sub-graph $26$) or high flow only between one pair of vertices (sub-graph $24$) than high flow between the three pairs of vertices (sub-graph $11$). In the static network, on the other hand, the high flow triangle (sub-graph $11$) is more represented in both networks, but still, not as much as the low flow triangles (sub-graphs $26$ and $24$). In the case of triplets, the symmetric sub-graphs are significantly anti-represented (sub-graphs $5$ and $10$). The symmetric high flow sub-graphs is not significantly represented (sub-graph $1$). This anti-representativity of symmetric triplets is also observed in directed networks of online social networks and web-pages~\cite{Milo04}.

\subsection{Higher order motifs}

Using the flow network defined in section~\ref{sec_flow}, it is possible to extract information about sub-graphs with any number of vertices. High order motifs provide detailed information of the flow within higher neighborhoods~\cite{Costa06} but the number of possible sub-graphs grows significantly for each added vertex. Furthermore, the computational cost makes the detection of higher order motifs impractical for networks of moderate size and link density such that other methods are more adequate to study the flow properties of large groups of vertices~\cite{Rosvall08, Kim10}. We therefore restrict our analysis to sub-graphs with $4$ vertices.

\begin{figure}[tbh]
\centering
\includegraphics[scale=0.38]{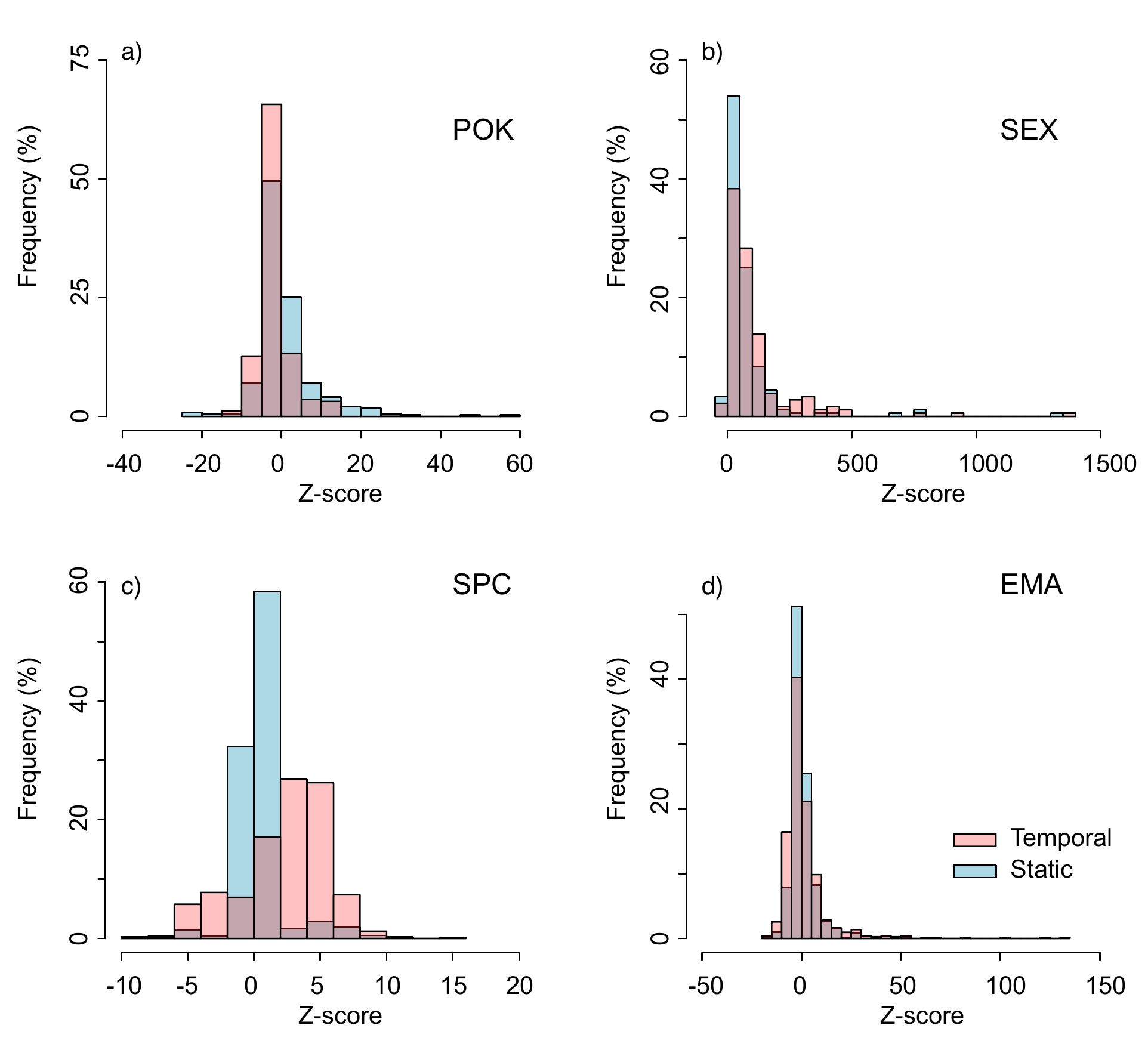}
\caption{Frequency of sub-graphs with a certain Z-score, for the temporal and static versions of the empirical networks. The results for SPM are qualitatively similar to EMA and thus omitted.}
\label{panel_7}
\end{figure}

We measure the Z-scores of the $4$-vertex sub-graphs by using $100$ realizations of the randomized version. Figure~\ref{panel_7} shows the frequency of sub-graphs with a certain value of Z-score for both temporal and static frameworks. In the static framework, sub-graphs typically have Z-scores concentrated around zero, while a larger number of sub-graphs have higher Z-scores in the temporal framework. These results indicate that the time order significantly constrains the flow such that more types of sub-graphs become representative whenever the temporal framework is adopted; these sub-graphs provide finer details about the relevance of the routes of spreading. In the SEX and EMA networks, there are highly represented motifs in both temporal and static frameworks, while in the POK and SPC cases, motifs are more distinguishable in the temporal and static frameworks respectively.

\begin{figure}[tbh]
\centering
\includegraphics[scale=1]{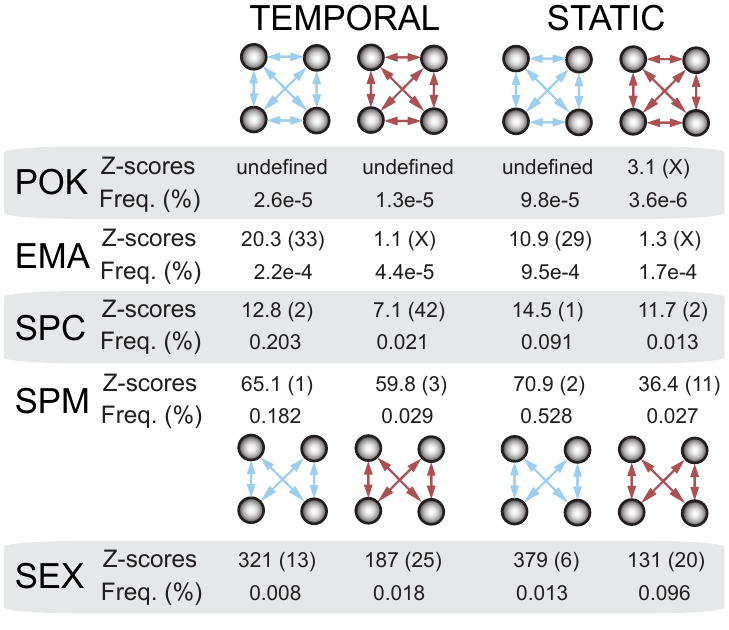}
\caption{The Z-score and the frequency of the fully connected $4$-vertex motifs where all links are either low-flow or high-flow, for the temporal and static frameworks. The number in parenthesis indicates the rank of the sub-graph in terms of its representativity (Z-score) in respect to the other sub-graphs of the same network and framework. The SEX case is the only purely bipartite network, which means that connections between different classes of vertices are non existent.}
\label{panel_8}
\end{figure}

Rather than looking to several individual sub-graphs or motifs, we turn our attention to a special structure, the fully connected 4-vertex sub-graph. By measuring both the Z-scores and the relative frequency, we identify that the low-flow (and not the high-flow) sub-graphs are the most significant in all studied datasets (Fig.~\ref{panel_8}). The relative frequency of low-flow sub-graphs is typically one order of magnitude higher than the high-flow cases on both temporal and static frameworks. With few exceptions, the Z-scores are also significantly higher for the low-flow sub-graphs. In the SEX network, which is bipartite, high-flow sub-graphs are more frequent but the Z-scores are significantly higher for the low-flow case as well. In accordance with the results for 3-vertex sub-graphs, these symmetric low- and high-flow sub-graphs are also more represented in the static versions of the empirical networks, reinforcing the conclusions that time order reduces the availability of certain paths and consequently increases the variability of flow on adjacent vertices.

The results suggest that single structures (i.e. cliques of 3 and 4 vertices) are insufficient to properly quantify the flow, or the ability to propagate information or infections, within adjacent vertices. In fact, in static networks, the spreading processes are constrained by the connectivity of the vertices linked to these single structures, i.e. the flow within a structure depends on the context in which the structure is placed. In the temporal framework, on the other hand, not only the topological context but also the time sequence of the links, within and outside the structure, play a role to define possible routes of propagation and thus regulate the flow between adjacent vertices.

 \section{Conclusions}

The increasing availability of high-resolution temporal network data has encouraged researchers to study the effects of temporal constrains in spreading processes on networks. While several studies suggest that temporal constrains regulate the speed of the spreading processes, it is unclear how these constrains affect the spreading between vertices. To address this problem, we have proposed a methodology to quantify the limits of flow between adjacent vertices by using a random walker dynamics and motif analysis. The normalized frequency of hops between two vertices gives an estimate of the overall flow between these vertices. By labeling the links into high and low-flow classes, we convert the original undirected network into a more complex network (with two types of links and directions) containing information about the dynamics. We introduce the concept of flow-motifs as particular sub-graphs observed significantly more often than one expects by chance.

We compare the abundance of $3$-vertex sub-graphs in the temporal networks in respect to the static case and identify that in the case of emails exchange (and face-to-face interactions), the regularity of contacts and persistence of vertices results on little change in the limits of flow whenever the static or temporal framework is used. On the other hand, in the case of communication between members of a dating site (and of a sexual network), the flow between adjacent vertices changes significantly in the temporal framework such that the static approximation poorly represents the structure of contacts. This happens because in this category of networks, the vertex turnover is high and contacts vary between active members such that links active at different times are mixed in the aggregated static structure. We have also observed that triangles of connected vertices do not necessarily correspond to high flow between the vertices forming the triangular structure. In fact, for the temporal network, the most representative triangle sub-graph corresponds to symmetric low flow between the three vertices and the second most representative triangle to high flow between a pair of vertices and low flow in the other two links. Similar results are obtained when using $4$-vertex sub-graphs. In that case, on both temporal and static frameworks, sub-graphs with all low-flow links are more often and more representative (higher Z-score) than sub-graphs with all high-flow links. These results suggest that low order fully connected structures (cliques) do not necessarily imply on high flow between the associated vertices. The flow, and possibly other spreading processes, within a structure depends not only on the topological context in which the structure is placed but also in the time order that links are established. In the studied datasets, the static framework typically overestimates the importance of low order cliques (e.g. the triangle).

The proposed methodology is general and can be applied to study other processes, as for example, epidemics, opinion dynamics, or traffic. It provides a summarized description of the potential of communication, influence, or transmission between small groups of vertices. This description can be used to identify vertices and links relevant to regulate spreading processes not only in the context of temporal networks but also in static networks.

\acknowledgments

The authors thank Martin Rosvall for enlighten discussions, Taro Takaguchi for pointing out missing references, Petter Holme and the SocioPattern consortium for sharing, respectively, the dating community and the face-to-face contacts datasets. LECR is beneficiary of a FSR incoming post-doctoral fellowship of the Academie universitaire Louvain, co-funded by the Marie Curie Actions of the European Comission. Computational resources have been provided by the supercomputing facilities of the Universit\'e catholique de Louvain (CISM/UCL) and the Consortium des \'Equipements de Calcul Intensif en F\'ed\'eration Wallonie Bruxelles (CECI) funded by FRS-FNRS.

\bibliographystyle{elsarticle-num}
\bibliography{Rocha_motifs}
\end{document}